# Stress shielding at the bone-implant interface: influence of surface roughness and of the bone-implant contact ratio

Maria Letizia Raffa[a], Vu-Hieu Nguyen[a], Philippe Hernigou[b,c], Charles-Henri Flouzat-Lachaniette[b,c] Guillaume Haiat[a,*]

[a] CNRS, Laboratoire Modélisation et Simulation Multi Echelle, MSME UMR 8208, CNRS, UPEC, F-94010, Créteil, France

[b] Service de Chirurgie Orthopédique et Traumatologique, Hôpital Henri Mondor AP-HP, CHU Paris 12, Université Paris-Est, 51 avenue du Maréchal de Lattre de Tassigny, 94000 Créteil, France.

[c] INSERM U955, IMRB Université Paris-Est, 51 avenue du Maréchal de Lattre de Tassigny, 94000 Créteil, France.

[*]**Corresponding author:** Guillaume Haiat; *Tel*: (+33) 1 45 17 14 41; *Fax*: (+33) 1 45 17 14 33; *E-mail*: guillaume.haiat@univ-paris-est.fr

**Running title:** Stress shielding at the bone-implant interface

**Authors' contributions**

MLR performed the computation and wrote the first draft. VHN worked on the computation and the study design. PH and CHFL critically revised the work. GH supervised the work. All authors read and approved the final manuscript.

**Abstract**

Short and long-term stabilities of cementless implants are strongly determined by the interfacial load transfer between implants and bone tissue. Stress-shielding effects arise from shear stresses due to the difference of material properties between bone and the implant. It remains difficult to measure the stress field in periprosthetic bone tissue. This study proposes to investigate the dependence of the stress field in periprosthetic bone tissue on i) the implant surface roughness, ii) material properties of bone and of the implant, iii) the bone-implant contact ratio. To do so, a microscale 2-D finite element model of an osseointegrated bone-implant interface was developed where the surface roughness was modeled by a sinusoidal surface. The results show that the isostatic pressure is not affected by the presence of the bone-implant interface while shear stresses arise due to the combined effects of a geometrical singularity (for low surface roughness) and of shear stresses at the bone-implant interface (for high surface roughness). Stress-shielding effects are likely to be more important when the bone-implant contact ratio value is low, which corresponds to a case of relatively low implant stability. Shear stress reach a maximum value at a distance from the interface comprised between 0 and 0.1 time roughness wavelength $\lambda$ and tend to 0 at a distance from the implant surface higher than $\lambda$, independently from bone-implant contact ratio and waviness ratio. A comparison with an analytical model allows validating the numerical results. Future work should use the present approach to model osseointegration phenomena.

# 1. Introduction

Cementless implants have been used clinically for more than forty years, leading to important progresses in orthopedic surgery. However, despite their routine clinical use, short and long-term implant failures still occur and remain difficult to anticipate. After surgery, the high contrast of mechanical properties between periprosthetic bone tissue and the implant biomaterial may causes bone resorption[1], a phenomenon defined as "stress-shielding" that is visible radiographically[2]. Stress-shielding may cause aseptic loosening and a loss of bone mineral density that increases the risk of periprosthetic fracture and makes revision surgery more difficult[3]. Although some studies on total hip arthroplasty found that stress-shielding may also have no clinical consequences[2], even at a very-long term with anatomic hydroxyapatite femoral components with high wear rates[4], monitoring and preventing stress-shielding has been recognized as an important action to obtain long-term fixation of joint arthroplasties[3].

Different authors have developed new biomaterials with lower mechanical properties compared to titanium alloy, leading to a reduction of stress-shielding related effects[5,6].

Long-term stability (also referred to as "secondary" stability) is obtained after the healing processes corresponding to osseointegration phenomena[7]. The implant primary stability is reinforced by the creation and maturation of newly formed bone tissue around the bone-implant interface (BII)[8,9]. Osseointegration phenomena are stimulated by stresses applied to bone tissue through the BII and the bone microstructure has been shown to be aligned along the principal local loading direction, suggesting that mechanical stimuli act as local regulators[10]. The quality of osseointegration phenomena is strongly influenced by the implant surface roughness[11], which can be obtained using different processes such as sand blasting[12], plasma spraying[13] or laser blasting[14]. Interfacial load transfer between implants and bone tissue determines osseointegration phenomena. Many biomechanical factors, including i) external loading, ii) implant material properties, geometry and surface properties and iii) periprosthetic bone tissue quality and quantity (defined by the bone-implant contact ratio, denoted as BIC) strongly affect stress and strain fields distribution around osseointegrated implants. All these intercorrelated features have a multiscale (from the nanometric scale up to the organ scale) and multitime (from seconds up to several months) nature and contribute to the evolution of the biomechanical properties of the BII. As a consequence, the mechanical and microstructural properties of bone tissue around the implant surface (at a distance of up to around 200µm from the implant) are determinant for the evolution of the implant stability as well as for the surgical success[15].

Despite the development of experimental techniques based on acoustical methods [10,16–18] to characterize the BII properties, it remains difficult to measure the biomechanical properties of the BII *in vivo* and there is a lack of experimental data at the scale of 1 to 100 µm[19]. In this context, finite element (FE) modeling represent an interesting modality because all parameters can be controlled and investigated independently, which is not possible when following an experimental approach. Numerical simulations have been extensively used to investigate the stress field in bone tissue at the scale of the implant (macroscale)[20–25], allowing to investigate the influence of implant design and mechanical properties on the implant stability. Nevertheless, the effects of changes of micromechanical properties (such as the BIC ratio, surrounding bone properties and the implant surface roughness) on the spatial variation of the stress field around the BII remain unclear.

The aim of this study is to investigate the dependence of the local stress field in bone tissue located around the BII on the implant mechanical properties and environment. To do so, a microscale 2-D FE model based on previous work by authors[26] is proposed. The model accounts for the effects of implant surface roughness, of the bone and implant stiffness and of the BIC ratio on the local stress field in bone tissue under tensile loading. The spatial variation of the stress field in the direction perpendicular to the implant interface is also investigated.

## 2. Material and methods

### 2.1. Mechanical model

#### 2.1.1. Geometry

A 2-D micromechanical model of the contacting region between a cementless implant and bone tissue is proposed in what follows. The model comprises two half-spaces corresponding to an implant and to bone tissue respectively, separated each other by an irregular interphase (i.e. a zone comprising the implant surface, bone tissue in partial contact with the implant and a region filled with void in the simulation), as shown in Fig.1.

The implant surface in contact with bone tissue was assumed to be rough and was modeled by a sinusoidal description, similarly as what was done in previous studies [27–29]. The one-dimensional sinusoidal surface of the implant of amplitude $2\Delta$ and wavelength $\lambda$ was geometrically defined as:

$$F(x) = \Delta \left[1 - \cos\left(\frac{2\pi x}{\lambda}\right)\right], \qquad x \in \left[0, \frac{\lambda}{2}\right] \tag{1}$$

where $F$ is the coordinate of the implant surface as a function of the local abscissa $x$, as shown in Fig. 1c. Note that the standard parameters used in surface engineering to describe a surface roughness (*i.e.* the arithmetical mean roughness $R_a$ and the mean spacing $S_m$) can be related to $\Delta$ and $\lambda$ by : $R_a \cong \frac{2\Delta}{\pi}$ and $S_m \cong \lambda$. The waviness ratio $\Delta/\lambda$ was assumed as a model parameter varying between 0.01 and 0.5. We verified that computations realized with different values of $\Delta$ and $\lambda$ but identical values of waviness ratio $\Delta/\lambda$ gave the same results, which justifies the choice of the waviness ratio as a suitable parameter of the propsed model. The lowest limit (0.01) of the waviness ratio represents a "microscopic" roughness corresponding to an implant surface roughness obtained by sandblasting and/or acid etching, while the highest limit corresponds to a "macroscopic" roughness due to the implant threading, for example in the case of the pedicle screws[20]. The value 0.1 was taken as the reference value for $\Delta/\lambda$ in what follows.

Due to the symmetry of the proposed configuration, a region of interest (ROI) comprising a half period $\lambda/2$ of the BII (see Fig. 1b,c) was considered. The origin of the reference system was defined at the point $O$ shown in Fig. 1c. The ROI shown in Fig. 1c comprised bone tissue and implant domains having a same thickness $H = 1$ mm along the $y$-direction.

The BIC ratio, which corresponds to the ratio of the bone in direct contact with the implant surface, was described by the parameter $h$ defined in Fig. 1c. Due to the sinusoidal geometry of the implant roughness, the BIC ratio can be obtained geometrically as follows:

$$\text{BIC} = \frac{L_P}{L_T} = 1 - \frac{E\left(\frac{2\pi x_T}{\lambda}\middle| -4\pi^2 \left(\frac{\Delta}{\lambda}\right)^2\right)}{2E\left(-4\pi^2\left(\frac{\Delta}{\lambda}\right)^2\right)}, \qquad (2)$$

where $L_P$ and $L_T$ represent the arc length of the sinusoidal implant surface in contact with bone tissue and the total arc length of the implant boundary, respectively. $L_P$ and $L_T$ depend on the amplitude $\Delta$ and the wavelength $\lambda$ of the sinusoidal implant surface through the operators $E(z) = E\left(\frac{\pi}{2}\middle|z\right)$ and $E = (z|m) = \int_0^z \sqrt{1 - m\sin^2(t)}\, dt$, which represent the complete and incomplete elliptic integral of the second kind, respectively. In Eq. (2), $x_T$ corresponds to the abscissa of the point T, as indicated in Fig. 1c.

The BIC ratio is a parameter of the proposed model and it assumed to vary between 5% and 80%[15]. For each BIC, Δ and λ values the parameter $h$ (see Fig. 1c) was calculated following this two-step procedure:

i. the value of abscissa $x_T$ was explicitly derived from Eq. (2);

ii. the obtained abscissa was injected in Eq. (1) to calculate the corresponding value of $h$.

*2.1.2. Materials*

All materials were assumed to be linear-elastic and to have homogeneous isotropic mechanical properties. Three different implant materials used in orthopedic applications were considered: i) a titanium alloy (Ti-6Al-4V) with a Young's modulus $E$ of 113 GPa[30]; ii) a Ti-Nb-Zr alloy (Ti-28Nb-35.4Zr) with $E = 51$ GPa [31] and iii) a metal-polymer composite (Ti-35BPA) with $E = 4.4$ GPa [32]. Two values of the bone Young's modulus $E_b$ were tested in order to simulate i) cortical bone tissue: $E_b = 2$ GPa [33] and ii) trabecular bone: $E_b = 0.2$ GPa[34]. All materials had a Poisson ratio $v$ equal to 0.3.

*2.1.3. Boundary conditions and assumptions*

The boundaries conditions, represented in Fig. 1c, were symmetric ($u_x = 0$) on the boundaries parallel to the $y$-axis. The boundary condition was fixed ($u_y = 0$) on the lower boundary ($y = -H$) of the bone domain. A uniaxial normal tension $\sigma_0 = 25$ MPa was applied on the upper implant boundary ($y = 2\Delta+H$). Continuity in terms of displacement and normal stress fields holds at the contacting surfaces between bone and implant, which corresponds to a perfect contact condition. Since the out-of-plane dimension of the ROI was assumed to be larger than its height ($2H + 2\Delta$) and the applied load was homogeneous, the assumption of plane strain was made.

## 2.2. Numerical simulations

A standard linear elastic problem was solved by the finite element (FE) method. All numerical analyses were carried out using Comsol Multiphysics® (Stockholm, Sweden).

### 2.2.1. Mesh generation and convergence

The finite element mesh used to run the numerical simulation, which was slightly changed depending on the geometrical parameters, typically contained around 5700 second-order triangular Lagrange elements, leading to a global system with about 21700 degrees of freedom. The interpolation functions for the displacement field were quadratic.

A convergence study was performed to choose the suitable element size for all values of $\Delta/\lambda$ considered in the present study. Note that the mesh size had to be lower at the tip of the non-contact zone at the BII in order to better describe the stress concentration, which occurs especially for lower values of $\Delta/\lambda$. The minimum element size was set equal to $3\times10^{-5}$ m. With the chosen mesh size, the local error related to the displacement along *y* $u_y$, is $\frac{\|u_y^r - u_y^a\|_{L^2}}{\|u_y^r\|_{L^2}} = 0.006\%$, where $\|\bullet\|_{L^2}$ indicates the $L^2$ norm, $u_y^r$ and $u_y^a$ are respectively the reference displacement obtained with a finest mesh size and the approximated displacement calculated with the chosen mesh size (defined through the convergence study).

### 2.2.2. Parametric study and stress field indicators

Parametrical analyses were carried out in order to investigate the influence of the BIC ratio (between 5% and 80%) and of the waviness ratio $\Delta/\lambda$ (between 0.01 and 0.5) on the spatial variation of the stress field in periprosthetic bone tissue. For each configuration, two values of the bone Young's modulus ($E_b$ = 0.2 and 2 GPa) and three different implant materials with Young's modulus $E$ equal to 113 GPa, 51 GPa and 4.4 GPa were tested. The reference configuration corresponds to the following parameters: BIC = 50%, $\Delta/\lambda$ = 0.1, $E$ = 113 GPa (Ti alloy) and $E_b$ = 2 GPa (cortical bone).

In order to characterize the spatial variation of the stress fields in bone tissue, two parameters were chosen: the maximal shear stress $\tau_{max}$ and the isostatic pressure $p_{iso}$. Both parameters are associated to the principal stresses of the material as follows:

$$\tau_{\max}(x, \tilde{y}) = \frac{\sigma_{p1} - \sigma_{p3}}{2}, \tag{3}$$

$$p_{iso}(x, \tilde{y}) = \frac{\sigma_{p1} + \sigma_{p2} + \sigma_{p3}}{3}, \tag{4}$$

where $\tilde{y} = \frac{-y}{\lambda}$ is the non-dimensional coordinate indicating the distance from the implant surface (see Fig. 1c), $x$ is the local abscissa and $\sigma_{p1}$, $\sigma_{p2}$, $\sigma_{p3}$ were the first, second and third principal stresses in bone tissue, respectively (with $\sigma_{p1} > \sigma_{p2} > \sigma_{p3}$).

The mean value of $\tau_{max}$ and of $p_{iso}$ as a function of $x$ was then determined for each values of $\tilde{y}$ following:

$$\langle \tau_{max} \rangle(\tilde{y}) = \frac{2}{\lambda} \int_0^{\frac{\lambda}{2}} \tau_{max}(x, \tilde{y}) \, dx, \tag{5}$$

$$\langle p_{iso} \rangle(\tilde{y}) = \frac{2}{\lambda} \int_0^{\frac{\lambda}{2}} p_{iso}(x, \tilde{y}) \, dx. \tag{6}$$

The average values $\langle \tau_{max} \rangle(\tilde{y})$ and $\langle p_{iso} \rangle(\tilde{y})$ were used as indicators of the level of the stress field in periprosthetic bone tissue.

In order to compare the results obtained with different implant materials, the following parameters were defined:

$$\overline{\Delta \tau}_{Ti,Zr}(\tilde{y}) = \frac{2}{\lambda} \int_0^{\frac{\lambda}{2}} \left| \tau_{max}^{Ti}(x, \tilde{y}) - \tau_{max}^{Zr}(x, \tilde{y}) \right| dx, \tag{7}$$

$$\overline{\Delta \tau}_{Ti,BPA}(\tilde{y}) = \frac{2}{\lambda} \int_0^{\frac{\lambda}{2}} \left| \tau_{max}^{Ti}(x, \tilde{y}) - \tau_{max}^{BPA}(x, \tilde{y}) \right| dx, \tag{8}$$

where $\tau_{max}^{Ti}(x, \tilde{y})$ (respectively $\tau_{max}^{Zr}(x, \tilde{y})$ and $\tau_{max}^{BPA}(x, \tilde{y})$) corresponds to the value of the maximum shear stress when considering titanium alloy (respectively Ti-Nb-Zr alloy and metal-polymer composite) as the biomaterial used for the implant (obtained by Eq. (3)). The indicator $\overline{\Delta\tau}_{Ti,Zr}(\tilde{y})$ (respectively $\overline{\Delta\tau}_{Ti,BPA}(\tilde{y})$) corresponds to the difference between the maximum shear stress in bone when the implant is made of titanium alloy and the maximum shear stress in bone when implant is made of Ti-Nb-Zr alloy (respectively metal-polymer composite) averaged along $x$ at each distance $\tilde{y}$ from the implant surface.

Note that $\overline{\Delta\tau}_{Ti,Zr}(\tilde{y})$ and $\overline{\Delta\tau}_{Ti,BPA}(\tilde{y})$ depend on the waviness ratio $\Delta/\lambda$ and are indicators of the reduction of the stress-shielding effect due to the modification of the implant material.

## 3. Results

### 3.1. Effect of the variation of the BIC ratio

$\langle p_{iso}\rangle/\sigma_0$ was shown to weakly vary as a function of $\tilde{y}$. For all values of the BIC and of the waviness ratio $\Delta/\lambda$, the maximum relative variation between $\langle p_{iso}\rangle/\sigma_0$ and 0.6191 was always lower than 0.07%, the value of $\langle p_{iso}\rangle/\sigma_0 = 0.6191$ corresponding to the solution obtained for the planar BII. Consequently, the variation of the isostatic pressure will not be studied in what follows and only the results in terms of the maximal shear stress will be considered.

Figure 2 shows the effect of the BIC on the variation of the normalized maximum shear stress $\tau_{max}/\sigma_0$ averaged over $x$ as a function of the normalized depth $\tilde{y}$. The waviness ratio, the implant and the bone Young's moduli are taken equal to the reference configuration: $\Delta/\lambda = 0.1$, $E = 113$ GPa and $E_b = 2$ GPa. The maximum shear stress decreases as a function of the BIC for all values of $\tilde{y}$. Moreover, the maximum value of the maximum shear stress is obtained for relatively low values of $\tilde{y}$ comprised between 0 (for high BIC values) and 0.1.

For all BIC values, $\tau_{max}/\sigma_0$ tends towards the same value $\frac{\tau_{max}^H}{\sigma_0} = 0.286$ when $\tilde{y} > 1$, which corresponds to the value of the maximal shear stress obtained with a planar BII.

### 3.2 Effect of the waviness ratio $\Delta/\lambda$

Figure 3 shows the spatial distribution of the maximal shear stress $\tau_{max}$ around the BII for a BIC ratio equal to 50% and three waviness ratios $\Delta/\lambda = 0.01, 0.1$, and $0.5$. A stress concentration is obtained at the intersection of the implant, bone tissue and of the void, which can be explained by the singularity present at the crack tip. The stress concentration effect is shown to be more important for lower values of $\Delta/\lambda$, which can be explained by the fact that the configuration becomes closer to a crack tip configuration compared to the cases of higher values of $\Delta/\lambda$.

Moreover, the value of the shear stress at the BII increases as a function of $\Delta/\lambda$, which may be explained by the fact that the normal of the interface moves from almost parallel to the $y$ axis for small values of $\Delta/\lambda$ to almost parallel to the $x$ axis for $\Delta/\lambda = 0.5$. These two competing phenomena (crack tip related effect and the influence of the normal of the BII) may explain the spatial variation of $\tau_{max}$.

Figure 4 shows the variation of $\tau_{max}(x, \tilde{y} = 0.2)$ as a function of $x/\lambda$ for three values of the waviness ratios $\Delta/\lambda = 0.01, 0.1$, and $0.5$ and for a BIC ratio equal to 50%. Note that plotting $\tau_{max}$ for $\tilde{y} = 0.2$ allows to be sufficiently far from the BII to avoid too strong effects of the crack tip, and sufficiently close from the BII to obtain a significant spatial variation of $\tau_{max}$. For $\Delta/\lambda = 0.01$, the maximum value of $\tau_{max}$ is obtained for $x/\lambda = 0.2$, which can be explained by the influence of the stress concentration at the crack tip located at $x/\lambda = 0.25$. However, for $\Delta/\lambda \geq 0.1$, the maximum value of $\tau_{max}$ is obtained for $x = 0$ due to the stronger influence of the BII on the spatial distribution of shear stresses (see above and Fig. 3).

Figure 5 shows the variation of the normalized maximal shear stress $\tau_{max}/\sigma_0$ averaged over $x$ as a function of $\tilde{y}$ for different values of $\Delta/\lambda$ and of the BIC. Figure 5 shows that the maximal value of $\langle \frac{\tau_{max}}{\sigma_0} \rangle$ decreases when the BIC increases. The results obtained for the different values of the waviness ratio when BIC = 5% are qualitatively similar, while the effect of the waviness ratio on the distribution of the maximum shear stresses is significant for higher values of the BIC ratio.

### 3.3 Effect of the bone Young's modulus $E_b$

As indicated in Section 2.1, two bone types have been considered in the present study, i.e. cortical bone tissue (with $E_b = 2$ GPa) and trabecular bone tissue (with $E_b = 0.2$ GPa). Simulations considering both types of bone tissues were carried out for all values of BIC (between 5 and 80 %) and waviness ratio (between 0.01 and 0.5). The relative difference between the maximal shear stresses obtained with trabecular and cortical bone tissues is always lower than 1.9 %, which indicates that the bone material properties weakly affect the stress field distribution.

### 3.4 Effect of the implant stiffness

Figure 6 shows the variation of $\overline{\Delta\tau}_{Ti,Zr}(\tilde{y})$ and $\overline{\Delta\tau}_{Ti,BPA}(\tilde{y})$ corresponding to the average relative variation of the shear stress as a function of the waviness ratio $\Delta/\lambda$ for different values of $\tilde{y}$ and for a BIC ratio equal to 50%. The results show that the reduction of maximal shear stresses in bone tissue obtained with an implant made of Ti-Nb-Zr instead of Ti-6Al-4V is lower than that obtained with an implant made Ti-35BPA instead of Ti-6Al-4V. Moreover, $\overline{\Delta\tau}_{Ti,Zr}$ and $\overline{\Delta\tau}_{Ti,BPA}$ decrease as a function of $\tilde{y}$ and increase as a function of $\Delta/\lambda$.

## 4   Discussion

A usual assumption made in implant biomechanics consists in considering that osseointegration phenomena mainly occur in a region of interest located at a distance lower than around 100-200μm[20]. Nevertheless, the distribution of the stress field around the BII at the scale of several hundred micrometers remains unclear due to a lack of experimental evidences. The originality of the present study is to investigate the spatial distribution of the local stress field in the periprosthetic bone tissue, which is an important determinant of the implant stability. To this aim, a microscale 2-D FE model of an osseointegrated BII taking into account the effects of the implant roughness, of bone and implant stiffnesses and of the BIC was developed. The influence of these parameters on the spatial variation of the stress field around the BII is investigated, which leads to an estimation of the region of interest where the presence of the implant influences the stress field and where osseointegration is thus likely to be affected by the implant surface roughness.

The stress fields in bone tissue has been characterized in terms of maximal shear stress $\tau_{max}$ and isostatic pressure $p_{iso}$ associated to the principal stresses. The results obtained in the present study show that the isostatic pressure is not influenced by the implant surface roughness for all values of the waviness ratio and of BIC, which may be explained by the fact that $p_{iso}$ is related to the changes of the ROI volume that does not occur for the proposed model. However, the maximum shear stress is shown to be sensitive to the implant surface roughness. These results are consistent with the work by Anderson et al.[35] showing that a compressive loading condition acting on an elastic body with a wavy interface may induce a local state of shear stresses along the interface, even if a fully bounded interface is considered[35].

Figures 2 and 5 show the effect of the BIC on the maximal shear stress. The results show that stress-shielding effects are likely to be more important when the BIC value is low, which corresponds to a case of relatively low implant stability. These results emphasize the importance of maximizing the BIC ratio in order to maximize the chances of surgical success because of three reasons. First, increasing the BIC ratio is known to improve the implant stability, which is a strong determinant of implant success [45]. Second, increasing the BIC ratio is known to favor osseointegration phenomena, which are defined by multiscale and multi-time phenomena leading to bone apposition around the implant surface. Based on the results obtained herein, it seems that bone ingrowth (*i.e.* bone apposition at the implant surface at the scale of the surface roughness) is more important compared to bone ongrowth (*i.e.* bone apposition on the implant surface at the scale of the implant, without considering the surface roughness). Third, as shown in the present study, stress-shielding effects, which are detrimental to the implant success, are more important when the BIC ratio decreases.

The maximum shear stress exhibits a maximum value comprised between $\tilde{y} = 0$ and $\tilde{y} = 0.1$ (according to the BIC and to the waviness ratio $\Delta/\lambda$) and then decreases as a function of $\tilde{y}$ and tends asymptotically to the value $\tau_{max}^H$ (see subsections 3.1 & 3.2). Note that for relatively high BIC values ($> 50\%$), the maximum of $\langle \frac{\tau_{max}}{\sigma_0} \rangle$ is always reached for $\tilde{y} = 0$ (i.e. at the implant surface) and $\langle \frac{\tau_{max}}{\sigma_0} \rangle$ is then a strictly decreasing function of $\tilde{y}$. This last result provide an estimation of the region of interest where stress-shielding are likely to be significant.

The aforementioned distribution of $\langle \frac{\tau_{max}}{\sigma_0} \rangle$ as a function of $\tilde{y}$ may be explained by considering an analytical model corresponding to the Boussinesq theory[36]. As illustrated in Fig. 7a, the contact pressure induced by the implant surface is modeled by a linear load applied on a line of length $B$ acting on a semi-infinite space corresponding to bone tissue, with an amplitude $q$ that is taken equal to the normal stress averaged over the contacting BII derived from FE model. The value of $B$ is equal to $x_T$ (see Fig. 1c). The maximum shear stress $\tau_{max}^a(\tilde{y})$ derived from this analytical model for $x = 0$ writes[36]:

$$\tau_{max}^a(\tilde{y}) = \frac{q \sin \alpha}{\pi}, \qquad (9)$$

where $\alpha = 2 \tan^{-1}\left(-\frac{B}{2\lambda\tilde{y}}\right)$. The comparison between the analytical solution and the numerical solution for BIC = 5% and $\Delta/\lambda = 0.1$ is shown in Fig. 7b. A good agreement is obtained for the amplitude and the position of the peak of the maximum shear stress. Note that the numerical solution does not tend to zero, as the analytical solution does, which is due to the confined boundary conditions assumed in the numerical model (see Fig. 1c). The slightly difference between analytical and the numerical results shown in Fig. 7b is due to the difference in terms of the distribution of the contact pressure, which is approximated to a linear distribution in the analytical case. A good agreement between analytical and numerical models is obtained also for other values of the BIC as can be appreciated by comparing Fig. 5 and Fig. 7c. The analytical model allows understanding the decrease of the peak value of $\langle \frac{\tau_{max}}{\sigma_0} \rangle$ when BIC increases (see Fig. 7c), which is due to an increasing contact area (contact length in 2-D case) between implant surface and bone tissue.

As shown in Figs. 3&4, the stress distribution around the BII strongly depends on the waviness ratio $\Delta/\lambda$, which may be explained by the fact that the singularity at the tip of the non-contact zone is stronger for lower values of $\Delta/\lambda$, as shown in Fig. 3. When $\Delta/\lambda = 0.5$, the maximal shear stress $\tau_{max}$ is almost uniformly distributed along the implant surface in contact with bone tissue, and the localization of the stresses at the tip is negligible. This last result is consistent with the "crack-like" behavior of the non-contact zone that is highlighted for waviness ratio lower than 0.03 [26].

As shown in Fig. 5, $\langle\frac{\tau_{max}}{\sigma_0}\rangle - \langle\frac{\tau_{max}^H}{\sigma_0}\rangle \approx 0$ when $\tilde{y} \geq 0.8$ for all BIC and waviness ratios, which indicates that the region in the surrounding bone tissue affected by the surface roughness is located at a distance lower than the amplitude of the implant roughness. This result is consistent with the belief that the evolution of the bone properties, at a distance of several hundred micrometers from the interface, are determinant for the implant success; since such distance corresponds to typical roughness wavelength used in many orthopedic implants[20].

The effects of changes of the implant stiffness on stress-shielding phenomena was investigated in this study. Stress-shielding phenomena come from shear stresses at the BII generated by the strong gap of mechanical properties between bone tissue and the implant. The results shown in Fig. 6 provide a quantitative indication of the decrease of stress-shielding phenomena when using Ti-Nb-Zr alloy and a metal-polymer composite (Ti-35BPA) compared to titanium implant (Ti-6Al-4V). As expected, it is found a stronger decrease of stress-shielding effects for Ti-35BPA compared to Ti-28Nb-35.4Zr for all waviness ratio at all depth $\tilde{y}$ (see Fig. 6). $\overline{\Delta\tau}_{Ti,Zr}$ and $\overline{\Delta\tau}_{Ti,BPA}$ decrease as a function of $\tilde{y}$, which is due to fact that the maximal shear stress decrease as a function of $\tilde{y}$ for all implant roughness and BIC values (see Figs. 2&5).

This study has several limitations. First, a sinusoidal description of the implant surface roughness is used, similarly as papers [26, 37]. This description constitutes a strong approximation and considering the real surface texture is likely to lead to different results. However, comparable approaches have already been developed in contact mechanics [27-29,38] and arbitrary irregular surface may be easily generated from a collection of sinusoidal surfaces, which should be done in the future. Second, bone material properties were assumed to be linearly elastic, homogeneous and isotropic and fluid-structure interactions were neglected, similarly as what was done in various previous FE-based numerical studies [22-25,39]. However, newly formed bone properties are heterogeneous[40] and viscoelastic, which was neglected herein. The assumption of bone homogeneity implies a simplified modeling of the bone microstructure, which may include cavities at various scales[41,42]. Third, it was not possible to perform an experimental validation due to the difficulty of measuring stress distribution around the implant. Fourth, adhesion phenomena at the BII[43, 44] that may be important in particular at early stage of osseointegration were not taken into account, which should be considered in the future.

**Conclusion**

This study proposes a model of the influence of the implant surface roughness, of the stiffness of bone and implant and of the BIC on the local stress field in periprosthetic bone tissue. The proposed FE model shows that the influence of the surface roughness is limited to a distance corresponding approximately to the wavelength of the surface roughness, which is consistent with previous results. The results obtained in the present study emphasize the importance of maximizing the bone-implant contact ratio at the microscopic scale not only to improve osseointegration phenomena, but also to minimize stress-shielding effects that are shown to occur in bone tissue mostly in region of interests located at a distance of the order of the surface roughness from the implant surface.

Based on the computation of the stress field around the implant, future works should focus on introducing bone remodeling at the local scale in order to model osseointegration phenomena and the evolution of the BII.

**Acknowledgements**

This project has received funding from the European Research Council (ERC) under the European Union's Horizon 2020 research and innovation program (grant agreement No 682001, project ERC Consolidator Grant 2015 *BoneImplant*).

**Conflict of Interest**: the authors declare that they have no conflict of interest.

**Figure legends**

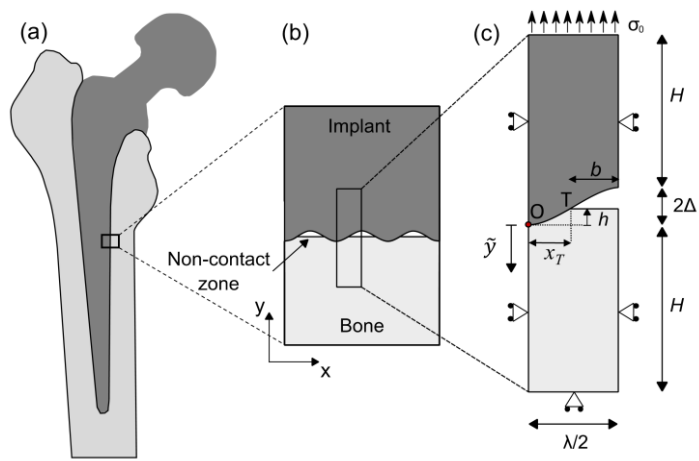

**Fig. 1** Geometrical configuration of the BII. (a): macroscopic description corresponding to a femoral stem taken as an example, (b): mesoscopic description of the BII and (c): microscopic description of the mechanical model of the region of interest.

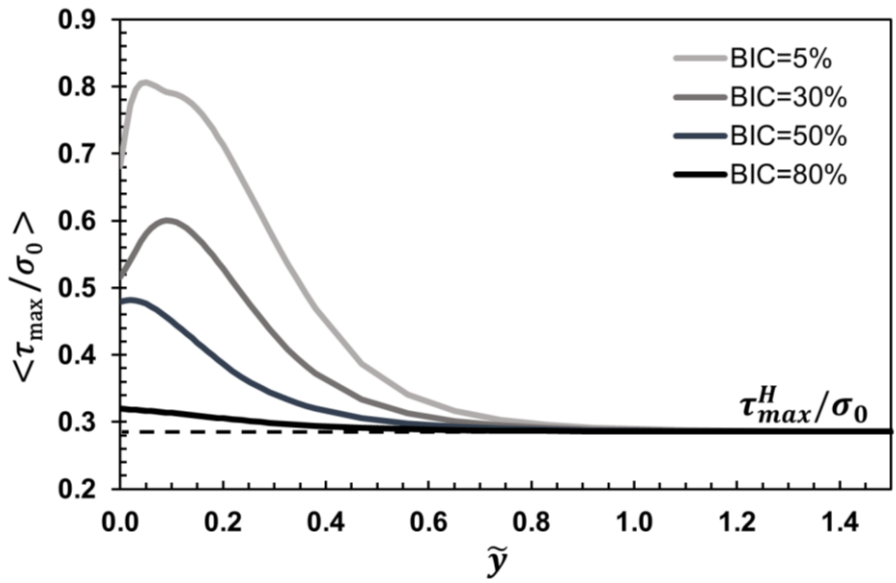

**Fig. 2** Variation of the normalized maximum shear stress $\tau_{max}/\sigma_0$ averaged over *x* as a function of the normalized depth for different values of the BIC. The waviness ratio and implant and bone Young's moduli are taken as: $\Delta/\lambda = 0.1$, $E = 113$ GPa and $E_b = 2$ GPa. The horizontal dashed line indicates the maximal shear stress obtained with a planar BII.

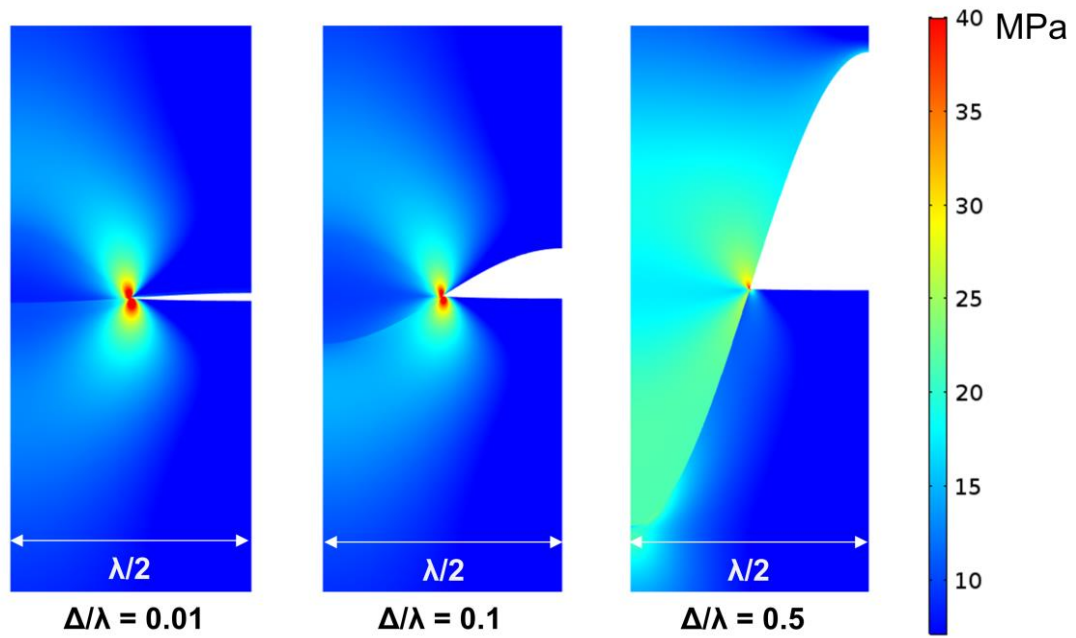

**Fig. 3** Spatial variation of the maximal shear stress $\tau_{max}$ (expressed in MPa) around the BII for BIC = 50% and three values of the waviness ratio $\Delta/\lambda$ = 0.01, 0.1, and 0.5. Implant and bone tissue Young's modulus are taken equal to $E$ = 113 GPa and $E_b$ = 2 GPa, respectively.

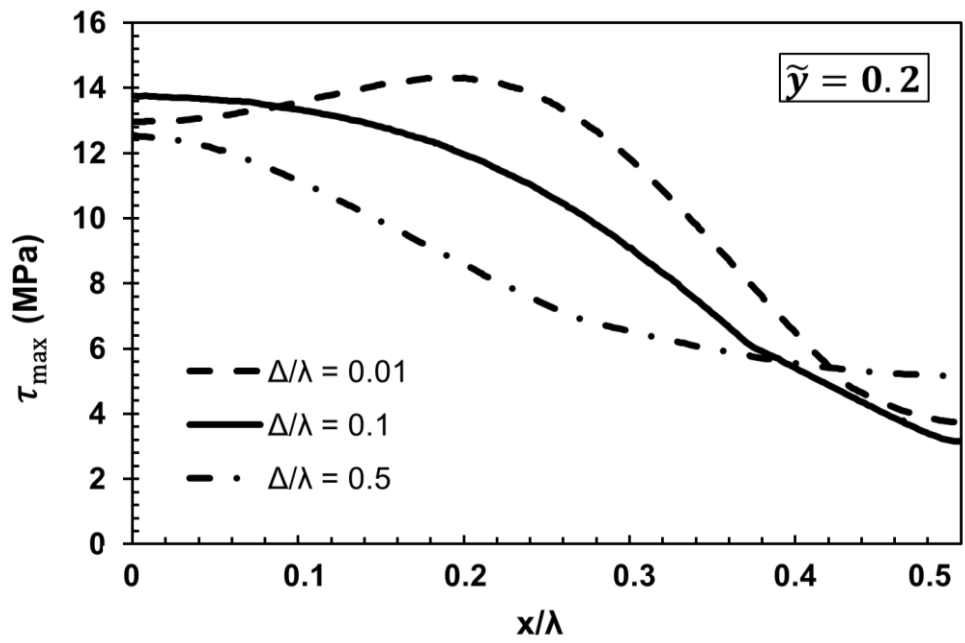

**Fig. 4** Variation of the maximal shear stress $\tau_{max}$ as a function of $x/\lambda$ for $\tilde{y} = 0.2$. Three values of the waviness ratio $\Delta/\lambda$ = 0.01, 0.1, and 0.5 are considered. The BIC ratio is equal to 50% and implant and bone Young's moduli are $E$ = 113 GPa and $E_b$ = 2 GPa, respectively.

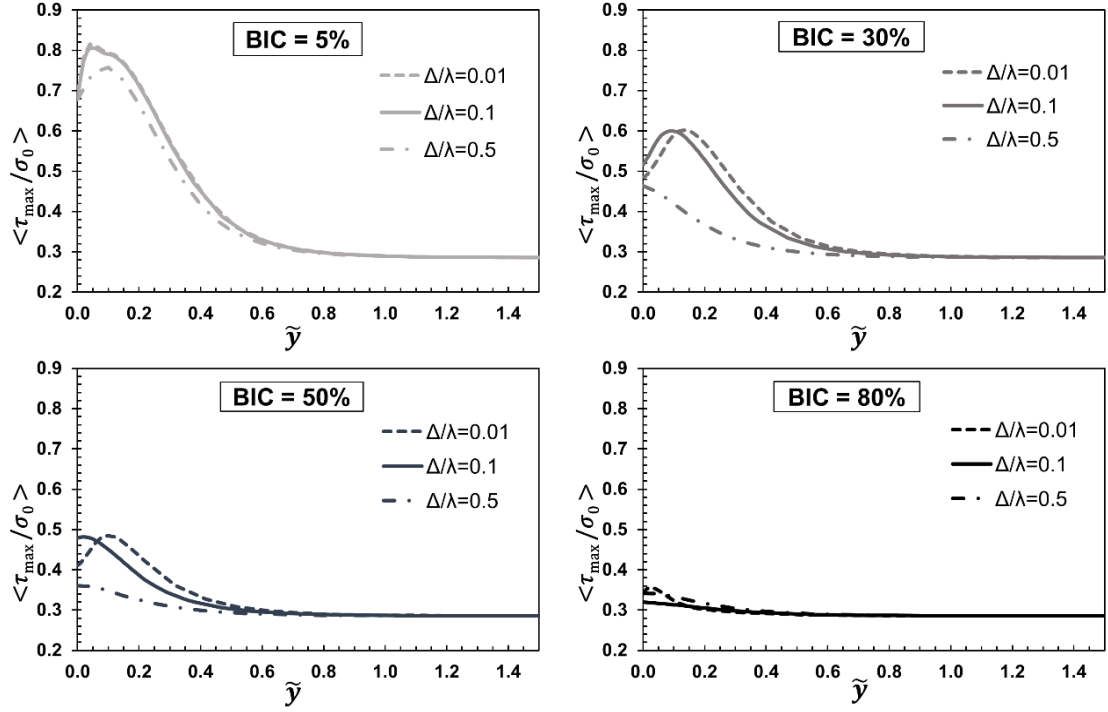

**Fig. 5** Variation of the normalized maximal shear stress $\tau_{max}/\sigma_0$ averaged over *x* as a function of $\tilde{y}$. Three values of the waviness ratio $\Delta/\lambda$ = 0.01, 0.1, and 0.5 and four values of the BIC ratio equal to 5, 30, 50, 80 % are considered, respectively. Implant and bone Young's moduli are equal to *E* = 113 GPa and $E_b$ = 2 GPa, respectively.

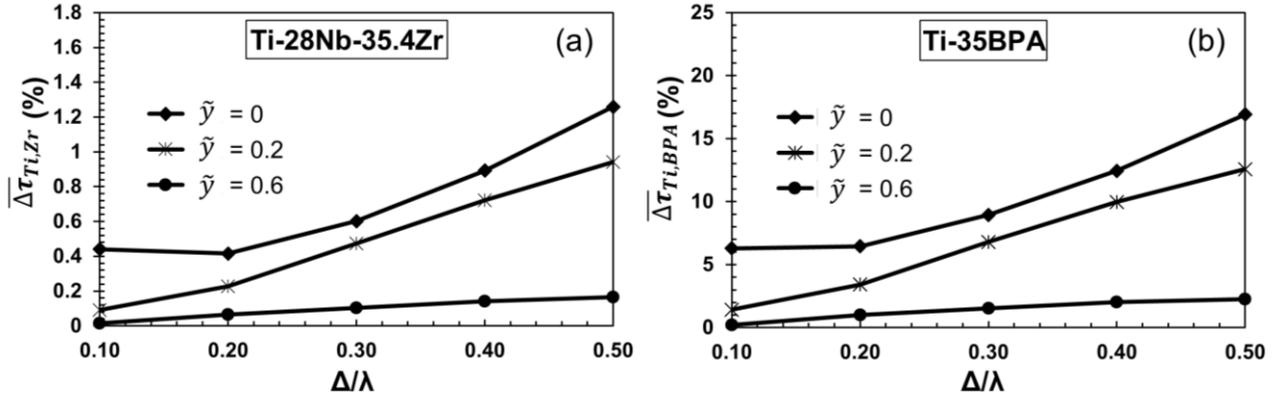

**Fig. 6** Effect of changes of implant biomaterial on the spatial distribution of the shear stress in bone tissue. Comparison of the stress field obtained between (a): Ti-6Al-4V and Ti-Nb-Zr alloy and (b): Ti-6Al-4V alloy and metal-polymer composite Ti-35BPA. The difference of the maximal shear stress $\overline{\Delta\tau}$ averaged as a function of $x$ is shown for waviness ratios $\Delta/\lambda = 0.1$, 0.2, 0.3, 0.4 and 0.5 and for $\tilde{y} = 0, 0.2, 0.6$. The BIC ratio is equal to 50% and the bone Young's modulus is $E_b = 2$ GPa.

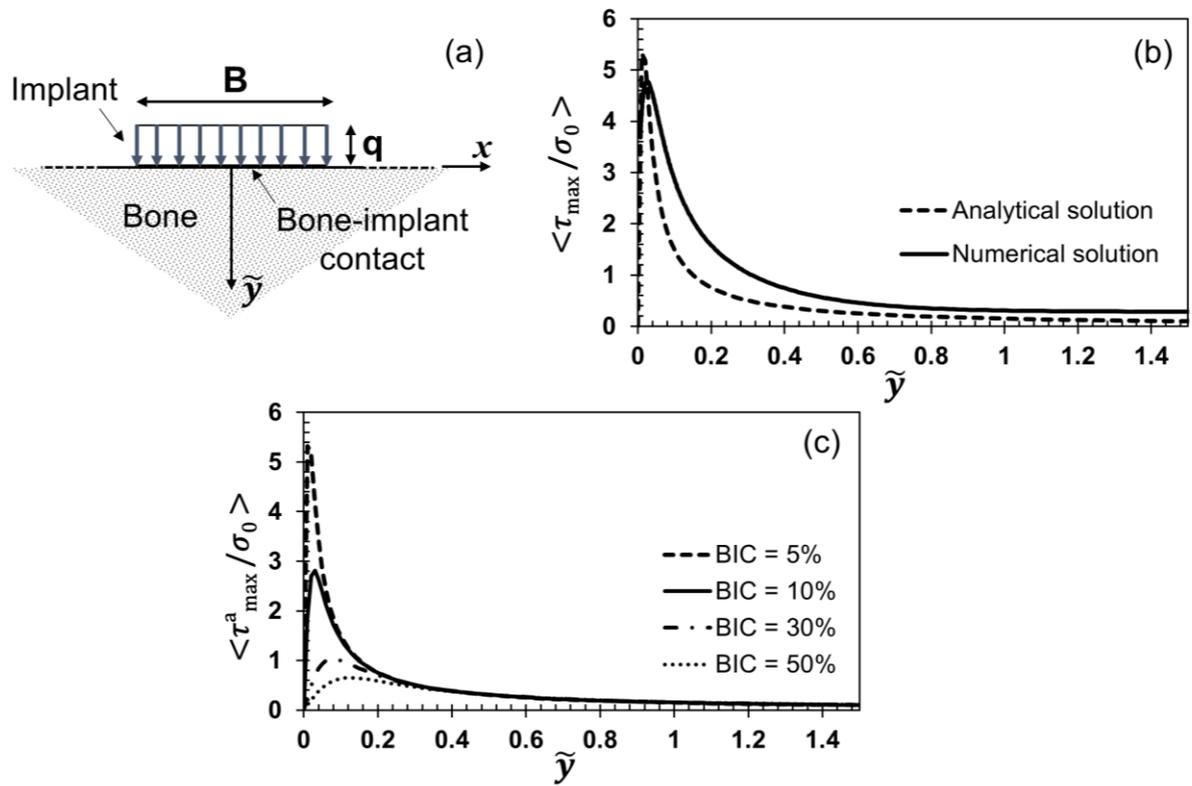

**Fig. 7** (a) Geometrical configuration corresponding to the Boussinesq theory; (b) comparison between the analytical and numerical results obtained for the maximum shear stress for BIC = 5% and $\Delta/\lambda = 0.1$; c) analytical results corresponding to the maximum shear stress $\tau^a_{max}/\sigma_0$ averaged over $x$ as a function of the normalized depth for different values of the BIC.